\documentclass[12pt]{article} 

\usepackage{epsfig}
\usepackage{amssymb}

\newcommand{\bra}{\langle}
\newcommand{\ket}{\rangle}
\newcommand{\half}{\frac{1}{2}}

\newcommand{\ep}{\epsilon}

\newcommand{\om}{\omega}  
\newcommand{\be}{\begin{equation}}
\newcommand{\ee}{\end{equation}}
\newcommand{\bea}{\begin{eqnarray}}
\newcommand{\eea}{\end{eqnarray}}
\newcommand{\bean}{\begin{eqnarray*}}
\newcommand{\eean}{\end{eqnarray*}}
\newcommand{\nn}{\nonumber}
\newcommand{\veck}{{\mathbf k}}
\newcommand{\vecp}{{\mathbf p}}

\newcommand{\vecx}{{\mathbf x}}

\newcommand{\vecnul}{{\mathbf 0}}
\newcommand{\omk}{\om_\veck}
\newcommand{\gak}{\gamma_\veck}

\newcommand{\pv}{{\mathbf p}}
\newcommand{\kv}{{\mathbf k}}

\newcommand{\pvuni}{\hat{{\mathbf p}}}

\newcommand{\puni}{\hat{p}}

\newcommand{\cP}{{\cal P}}

\newcommand{\qo}{\omega}
\renewcommand{\wp}{\omega_{\rm pl}}

\newcommand{\hm}{\hspace*{-0.6cm}}

\newcommand{\siml}{\lesssim}

\begin{document}

\title{
\vskip -100pt
{
\begin{normalsize}
\mbox{} \hfill hep-ph/0203177
\vskip  70pt
\end{normalsize}
}
Transport coefficients, spectral functions\\ 
and the lattice
\author{
Gert Aarts\thanks{
email: aarts@mps.ohio-state.edu}\addtocounter{footnote}{1} 
{} and 
J.\ M.\ Mart{\'\i}nez Resco\thanks{
email: marej@pacific.mps.ohio-state.edu}\addtocounter{footnote}{2}\\
{}\\
\normalsize
{\em Department of Physics, The Ohio State University}\\
\normalsize
{\em 174 West 18th Avenue, Columbus, OH 43210, USA}
}
}
\date{March 18, 2002}
\maketitle
 
\renewcommand{\abstractname}{\normalsize Abstract} 
\begin{abstract}
\normalsize 

Transport coefficients are determined by the slope of spectral functions
of composite operators at zero frequency. We study the spectral
function relevant for the shear viscosity for arbitrary frequencies in
weakly-coupled scalar and nonabelian gauge theories at high temperature 
and compute the corresponding correlator in euclidean time. We discuss 
whether nonperturbative values of transport coefficients can be extracted 
from euclidean lattice simulations.

\end{abstract}
 
\newpage

\section{Introduction}

In recent years there has been a growing interest in the calculation of
transport coefficients in quantum field theories at finite temperature.
With the advent of relativistic heavy-ion colliders, such as RHIC, a
proper knowledge of transport coefficients has become relevant since
hydrodynamical descriptions of heavy-ion collisions provide a useful tool
to analyse the experimental data \cite{Kolb:2001qz}. In practice the
extension of ideal relativistic hydrodynamics to include finite transport
coefficients is far from straightforward and applications to heavy-ion
physics have only just begun \cite{Muronga:2001zk}.

If the temperature is sufficiently high and the theory is weakly coupled,
transport coefficients can be computed in a perturbative expansion,
employing either kinetic theory or field theory using Kubo formulas.  It
turns out that the latter approach requires the summation of an infinite
series of higher-order diagrams, known as ladder diagrams, which has been
a serious drawback for its use. For a scalar theory, the higher-order
contributions in the loop expansion have been identified and summed in
Ref.\ \cite{Jeon:if}, using an intricate diagrammatic analysis, and the
leading-order results for the shear and bulk viscosities have been found.  
In Ref.~\cite{Jeon:1995zm} the equivalence of an effective Boltzmann
equation and the field-theoretical calculation is shown. Using a more
transparent analysis, the diagrammatic conclusions of Ref.\ \cite{Jeon:if}
have been confirmed recently \cite{Wang:1999gv}. So far, the only other
transport coefficient for which the ladder series has been summed
explicitly is the color conductivity in QCD~\cite{MartinezResco:2000pz}.
The viscosities in the scalar theory and color conductivity in QCD have
the property that the one-loop contribution and the ladder contributions
are of the same order in the coupling constant. However, for other
transport coefficients in gauge theories, such as the shear viscosity or
the electrical conductivity, the ladder contributions are in fact larger
than the one-loop one~\cite{Pisarski:1992hy}. Only recently and using
kinetic theory, a complete computation of the leading logarithmic order of
these transport coefficients has appeared~\cite{Arnold:2000dr}.  
Unfortunately, a full leading-order computation is still lacking.  
Ref.~\cite{Arnold:2000dr} provides a useful guide to the literature.

Euclidean lattice simulations offer in principle the possibility to
compute transport coefficients completely nonperturbatively
\cite{Karsch:1986cq}. However, transport coefficients are determined by
the small frequency limit of zero-momentum spectral functions of
appropriate composite operators (such as components of the energy-momentum
tensor) and spectral functions or other real-time correlators are not
readily available on a euclidean lattice, although recent progress has
been made with the Maximal Entropy Method (MEM)
\cite{Karsch:2001uw,Asakawa:2000tr}.\footnote{Note that for classical
field theories at finite temperature spectral functions can be computed
nonperturbatively by numerical simulations directly in real time
\cite{Aarts:2001yx}.} For that reason it was proposed in Ref.\
\cite{Karsch:1986cq} to introduce instead an ansatz for the spectral
function and fit the result to the numerical data for the euclidean
correlator, employing a standard dispersion relation between these two.
This approach was pursued more recently in Refs.\
\cite{Nakamura:1996na,Nakamura:1997bh}.

Motivated by these studies, our goal in this paper is to compute the
spectral function relevant for the shear viscosity at high temperature in
weakly-coupled scalar (Sec.\ \ref{secscalar}) and nonabelian gauge
theories (Sec.\ \ref{secgauge}). In the Conclusions we compare our
findings with the analysis carried out so far in Refs.\
\cite{Karsch:1986cq,Nakamura:1996na,Nakamura:1997bh}. It is found that
the ansatz used in these papers is inadequate and we suggest a better
one.  We also point out a potential problem in the calculation of
spectral functions at very low frequencies ($\om\to 0$) from euclidean
lattice correlators using the MEM approach.

\section{Correlation functions}

We start with a summary of basic relations between transport coefficients,
spectral functions and euclidean correlators, using the shear viscosity as
an example. The relations presented in this section are quite general and
valid for arbitrary transport coefficients.

The shear viscosity can be defined from a Kubo relation as   
\be
\eta = \frac{1}{20}\lim_{\om\to 0} \frac{1}{w} \int d^4x\, e^{i\om t} 
\left\bra \left[ \pi_{kl}(t,\vecx), \pi_{kl}(0,\vecnul) \right] 
\right\ket,
\ee
with $\pi_{kl}$ the traceless part of the spatial energy-momentum tensor.
The brackets denote the equilibrium expectation value at temperature $T$. 
We define spectral functions of hermitian (composite) operators, such as  
$\pi_{kl}$, as the expectation value of the commutator,
\be
\label{eqspeccomm}
\rho_{\pi\pi}(x-y) = \bra [ \pi_{kl}(x), \pi_{kl}(y) ]\ket,
\ee
and in momentum-space
\be
\rho_{\pi\pi}(\om,\vecp) = \int d^4x\, e^{i\om t-i\vecp\cdot \vecx} 
\rho_{\pi\pi}(t,\vecx). 
\ee
Spectral functions obey basic symmetry relations\footnote{
When most manipulations take place in real-space, it can be
convenient to define spectral functions such that they are real instead of 
purely imaginary in real-space. Here we stick to the usual convention and 
spectral functions are real in momentum-space.} 
$\rho_{\pi\pi}^*(x) = -\rho_{\pi\pi}(x) = \rho_{\pi\pi}(-x)$ and
$\rho_{\pi\pi}^*(\om,\vecp) = \rho_{\pi\pi}(\om,\vecp) = -
\rho_{\pi\pi}(-\om,\vecp)$ as well as the positivity condition
$\om\rho_{\pi\pi}(\om,\vecp) \geq 0$. The shear 
viscosity is determined by the slope at zero frequency:
\be
\eta = \frac{1}{20} \frac{d}{d\om}\rho_{\pi\pi}(\om)\Big|_{\om=0},
\ee
where $\rho_{\pi\pi}(\om) = \rho_{\pi\pi}(\om, \vecnul)$.
Since the spectral function is odd, we consider from now on positive 
$\om$ only.

The euclidean correlator (at zero spatial momentum) is given by 
\be
G^E_{\pi\pi}(\tau) = \int d^3x\, 
\bra 
\pi_{kl}(\tau,\vecx)\pi_{kl}(0,\vecnul)\ket_E\;\;\;\;\;\;\;\;(\tau=it),
\ee
where the imaginary time $\tau \in [0,1/T]$ and $G^E_{\pi\pi}(1/T-\tau) = 
G^E_{\pi\pi}(\tau)$.
The euclidean correlator and the spectral function are related via an 
integral equation, originating from a dispersion relation, 
\be
\label{eqrel}
G^E_{\pi\pi}(\tau) = \int_0^\infty \frac{d\om}{2\pi}\, 
K(\tau,\om)\rho_{\pi\pi}(\om),
\ee
with the kernel
\be
K(\tau,\om) = e^{\om\tau} n(\om) + e^{-\om\tau} \left[1+n(\om)\right]
= e^{-\om\tau} +2n(\om)\cosh \om\tau,
\ee
obeying $K(\tau,\om)=-K(\tau,-\om) = K(1/T-\tau,\om)$. The Bose 
distribution is
\be
n(\om) = \frac{1}{\exp(\om/T)-1}.
\ee

The low-frequency part of the spectral function contains all information 
on the transport coefficient and its effect on the euclidean correlator 
can be estimated quite easily. When $\om \ll T$, the kernel can be 
expanded as
\be 
\label{eqexp}
K(\tau,\om) = \frac{2T}{\om}
+ \frac{\om}{T}\left[\frac{1}{6}-\tau T(1-\tau T)\right] 
+ {\cal O}(\om^3/T^3),
\ee 
and all except the first term are suppressed. As a consequence we find 
that the contribution to the euclidean correlator from low frequencies,
\be 
\label{eqlow}
G^{E,\rm low}_{\pi\pi}(\tau) = 2T \int_0^{\om_\Lambda} 
\frac{d\om}{2\pi}\,\frac{\rho_{\pi\pi}(\om)}{\om},
\ee
is independent of $\tau$. The frequency cutoff ${\om_\Lambda} \ll T$ is
introduced here to justify the expansion of the kernel. We conclude that
the dominant effect of the low-frequency region is a constant
$\tau$-independent contribution to the euclidean correlator.

\section{Scalar field}
\label{secscalar}

We consider a one-component massless scalar field with a quartic 
$\lambda\phi^4/4!$ interaction.\footnote{We assume that the temperature 
is sufficiently high such that a possible zero-temperature mass $m_0^2 \ll 
\lambda T^2$ plays no role.} 
The one-particle spectral function is 
\be
\rho(x-y) = \bra [\phi(x),\phi(y)] \ket = G^>(x-y) - G^<(x-y).
\ee
The one-particle Wightman functions, $G^>(x-y) = \bra \phi(x)\phi(y)\ket$ 
and $G^<(x-y) = \bra \phi(y)\phi(x)\ket$, are related to the one-particle 
spectral function via the KMS condition 
\be
\label{eqKMS}
G^>(p) = \left[n(p^0)+1\right]\rho(p),\;\;\;\;\;\;\;\;
G^<(p) = n(p^0)\rho(p).
\ee
The traceless part of the spatial energy-momentum tensor reads
\be
 \pi_{kl} = \partial_k\phi\partial_l\phi 
-\frac{1}{3}\delta_{kl}\partial_m\phi\partial_m\phi.
\ee
The lowest-order skeleton diagram that contributes to the spectral
function for the shear viscosity in Eq.\ (\ref{eqspeccomm})  follows from
simple Wick contraction,
\be
\bra [\phi^2(x),\phi^2(y) ]\ket = 2\left[ G^{>2}(x-y) - 
G^{<2}(x-y)\right],
\ee
and we find
\be
\label{eqrho}
\rho_{\pi\pi}(\om) =
\frac{4}{3} \int \frac{d^4k}{(2\pi)^4}\, (\veck\cdot\veck)^2 
n(k^0)\rho(k^0,\veck)\left[\rho(k^0+\om,\veck) - 
\rho(k^0-\om,\veck)\right],
\ee
where we used the KMS conditions (\ref{eqKMS}) and $-n(-\om) = n(\om)+1$.
The four $\veck$'s in the integrand arise from the derivatives in 
$\pi_{kl}$. 
In the remainder of this section we compute the one-loop
spectral function (\ref{eqrho}) as a function of the external frequency.

The spectral function $\rho_{\pi\pi}$ depends on the one-particle
spectral functions $\rho$, which contain the quasiparticle structure of 
the theory at finite temperature. We describe this in some detail
since similar, though more complicated, considerations appear in gauge
theories. For hard momenta $|\veck|\sim T$, excitations are on-shell with
energy $|\veck|$. For softer momenta screening effects become important
and hard thermal loop (HTL) resummation yields a temperature-dependent
plasmon mass, $m^2 = m^2_{\rm th}(1-3m_{\rm th}/\pi T+\ldots)$ with
$m^2_{\rm th} = \lambda T^2/24$ \cite{Parwani:1991gq}. Finally, collisions
in the plasma result in a finite (but narrow) momentum-dependent width
$2\gamma_\veck\ll m\ll T$, changing the one-particle spectral function
from an on-shell delta function to a Breit-Wigner spectral function (see
below) \cite{Wang:1995qf}.

\begin{figure}[t]
\centerline{\epsfig{figure=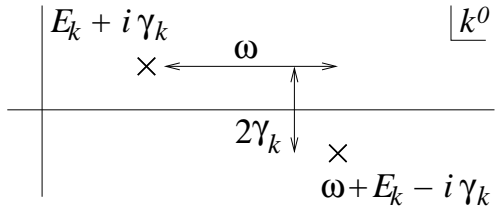,height=2.7cm}}
\caption{Typical configuration of poles in the complex $k^0$-plane for the 
evaluation of the one-loop spectral function $\rho_{\pi\pi}(\om)$, 
$E_\veck$ ($2\gamma_\veck$) denotes the quasiparticle energy (width).
}
\label{figpoles}
\end{figure}

We may now discuss the one-loop expression (\ref{eqrho}).
First we note that the integral is dominated by hard $\sim T$ momenta.
Therefore, for external frequencies that are not too small a simple 
on-shell delta function for the one-particle spectral functions suffices 
to find the dominant contribution. We will refer to this region as the 
high-frequency region. For smaller frequencies, however, the arguments of 
the delta functions come close, producing a so-called pinch singularity
\cite{Jeon:if}. The pinch singularity is screened by a finite external
frequency or width, whichever one is the largest. For very small
external frequencies, the inclusion of the width\footnote{And of ladder 
diagrams, see below.} is essential \cite{Jeon:if}. This situation is 
sketched in Fig.\ \ref{figpoles}. The
effect of these nearly pinching poles is to enhance the spectral function
compared to naive estimates. We will refer to the region where pinching 
poles are important as the low-frequency domain.

\begin{figure}[t]
\centerline{\epsfig{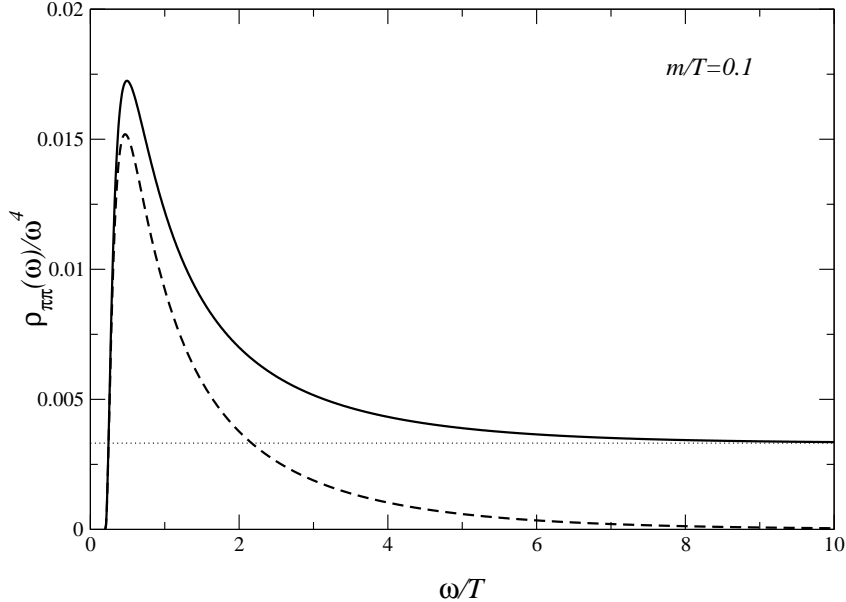}}
\caption{Contribution to the spectral function $\rho_{\pi\pi}(\om)/\om^4$ 
(full line) from decay/creation processes [see Eq.\ (\ref{eqhigh})] as a 
function of $\om/T$, with $m/T=0.1$. The contribution from the nearly 
pinching poles in the low-frequency region is discussed below. The dashed 
line shows the contribution proportional to the Bose distribution only. 
The dotted line indicates the asymptotic value.
}
\label{figabove}
\end{figure}

We start with the region where the frequency is much larger than the 
thermal width and no pinch-singularity problems are encountered. 
The one-particle spectral function can be taken on-shell and is
\be
\rho_0(k^0,\veck) = 2\pi\ep(k^0)\delta(k_0^2-\omk^2),
\ee
where $\omk = \sqrt{\veck^2+m^2}$ and $\ep(x)$ the sign-function. HTL 
effects are included in the mass parameter. 
Evaluating the integrals in Eq.\ (\ref{eqrho}) with the use of the 
delta-functions (the angular integrals are trivial) results in (recall 
that we take $\om>0$)
\be
\label{eqhigh}
\rho_{\pi\pi}(\om) = \theta(\om-2m)\frac{
\left(\om^2-4m^2\right)^{5/2}}{48\pi\om}\left[n(\om/2)+\half\right],
\ee
which is shown in Fig.\ \ref{figabove}. The physical processes are the
decay of a zero-momentum excitation with energy $\om$ into two on-shell
particles with equal and opposite momentum, and the inverse process of
creation. The decay process contributes also at zero temperature and makes
the spectral function increase as $\om^4$ at large frequencies. The
threshold at $\om=2m$ arises from (simple) HTL resummation. This concludes
the analysis of the spectral function in the high-frequency domain.

We continue with the low-frequency region where pinch singularities lead 
to a nontrivial enhancement of the spectral function. 
We replace the on-shell one-particle spectral functions with Breit-Wigner 
spectral functions:
\be
 \label{eqBW}
 \rho_{BW}(k^0,\veck) = \frac{1}{2\omega_{\veck}} 
 \left[ 
 \frac{2\gak}{(k^0 - \omk)^2 + \gak^2} -
 \frac{2\gak}{(k^0 + \omk)^2 + \gak^2} 
 \right].
\ee
The width $2\gamma_\veck$ is determined by the imaginary part 
of the retarded self energy
\be
\gamma_\veck = -\frac{\mbox{Im}\, \Sigma_R(\omk,\veck)}{2\omk},
\ee
and the dominant contribution at weak coupling arises from two-to-two 
scattering from the two-loop setting-sun diagram \cite{Jeon:if,Wang:1995qf}.
A convenient way to write this damping rate is as \cite{Wang:1995qg}
\be
\gamma_\veck =  \gamma \frac{T}{\omk} B(|\veck|/T;m/T),
\ee
where 
\be
\gamma = \frac{\lambda^2 T}{1536\pi},
\ee
determines the parametrical behaviour. 
The function $B$ contains the nontrivial momentum dependence and is
related to $A$ defined in Ref.\ \cite{Wang:1995qg} as
$B(|\veck|/T;m/T) = (6/\pi^2) A(|\veck|/T;m/T)$. In the limit of hard 
momenta and small mass \cite{Wang:1995qg}
\be
\lim_{m\to 0} \lim_{|\veck|\to \infty} B(|\veck|/T;m/T) = 1.
\ee
For analytical estimates we will neglect the momentum dependence and take
$B=1$. In the results obtained by numerical integration the full
momentum dependence is included.

We insert the Breit-Wigner functions into expression (\ref{eqrho}) for the
spectral function and perform the $k^0$ integral by integrating around the
poles in the complex plane.  We preserve only the dominant contributions
and discard all terms suppressed by (powers of) the coupling constant with respect
to the leading order contribution.  Breit-Wigner spectral functions have
four poles at complex energy-arguments $k^0 = \pm (\omk \pm i\gak)$.  From
the residue of these poles we keep $n(\omk \pm i\gak) \sim n(\omk)$.  The
Bose distribution $n(k^0)$ has poles along the imaginary axis at $k^0 =
2\pi i nT$, $n \in\mathbb{Z}$. However, the residues at these
poles are subdominant compared to those from the poles of the
Breit-Wigner functions. Hence we do not include these contributions. After
performing the $k^0$ integral we find
\be
\rho_{\pi\pi}(\om) = -\frac{4}{3} \int_\veck\, \frac{|\veck|^4}{2\omk}
\left\{ 
[n(\omk) - n(\omk-\om)] I(\om,\veck)\frac{\omk-\om}{\om^2+4\gak^2}
 - [\om \to -\om]
\right\},
\ee
with
\be
\int_\veck = \int \frac{d^3k}{(2\pi)^3},
\ee
and 
\be 
\label{eqI}
I(\om,\veck) = \frac{8\gak}{(\om-2\omk)^2+4\gak^2}.
\ee
For sufficiently large $\om$ and in the limit of small coupling (width)
we may take $I(\om,\veck) \to 4\pi\delta(\om-2\omk)$ and the result of 
the previous calculation in the high-frequency region is recovered, as it 
should. 

Let's now consider the low-frequency region, $\om \lesssim m$. Here we 
may approximate $I(\om,\veck) \simeq 2\gak/\omk^2$ and 
expand the difference between the Bose distributions to find
\be
\label{eqrhobelow}
\rho_{\pi\pi}(\om) = -\frac{8}{3} \int_\veck\, \frac{|\veck|^4}{\omk^2}
 n'(\omk) \frac{\om\gak}{\om^2+4\gak^2} 
\;\;\;\;\;\;\;\;(0\leq\om\lesssim m).
\ee
We note that the last factor controls the pinch singularity in precise
agreement with Fig.\ \ref{figpoles}. Although in principle 
$\rho_{\pi\pi}(\om)/T^4$ may depend on the three dimensionless
combinations $\om/T$, $\gamma/T$, and $m/T$, we find that in practice it 
only depends on $\om/\gamma$ and $m/T$. Since the integral is dominated by 
hard momenta the $m/T$ dependence is subdominant and the natural parameter 
on which the spectral function depends in this region is $\om/\gamma$. It is 
easy to see that the spectral function has a local maximum at 
$\om\sim\gamma$ and $\rho_{\pi\pi}(\om\sim\gamma)/T^4\sim 1$.

\begin{figure}
\centerline{\epsfig{figure=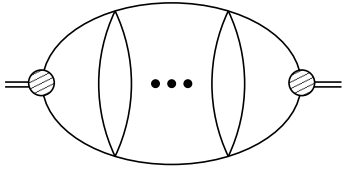,height=2.2cm}}
\caption{Ladder diagrams that contribute to $\rho_{\pi\pi}(\om)$ in the 
scalar theory. In the low-frequency region, $\om \lesssim \gamma$, 
the pinching-pole contributions from the ladder diagrams are equally 
important as the one-loop contribution. When $\om \gg \gamma$, 
pinching-pole contributions from ladder diagrams are suppressed.
}
\label{figscalarladder}
\end{figure}

For very small frequencies $\om\ll\gamma$ we expand and obtain
\be
\rho_{\pi\pi}(\om)/T^4 = a_1 \left(\frac{\om}{\gamma}\right) + 
\frac{a_3}{3!} \left(\frac{\om}{\gamma}\right)^3 + \ldots
\;\;\;\;(0\leq \om \ll \gamma),
\ee
with
\bea
&&a_1 = 
-\frac{2}{3T^4} \int_\veck\, \frac{|\veck|^4}{\omk^2}
 n'(\omk) \frac{\gamma}{\gak}  \simeq   \frac{5!\,\zeta(5)}{3\pi^2} \\
&&a_3 = 
\frac{1}{T^4} \int_\veck\, \frac{|\veck|^4}{\omk^2}
 n'(\omk) \left(\frac{\gamma}{\gak}\right)^3 \simeq   
-\frac{7!\,\zeta(7)}{2\pi^2},
\eea
where the $\simeq$ indicates that the final integrals are evaluated by 
neglecting the remaining momentum dependence of the damping rate (i.e.\ 
taking $B=1$)
as well as the thermal mass (the first approximation has numerically the 
largest effect). From this the one-loop viscosity follows as 
\be
\eta_{\rm 1-loop} = -\frac{1}{30}  \int_\veck\, \frac{|\veck|^4}{\omk^2}
 n'(\omk) \frac{1}{\gak} \simeq \frac{2\zeta(5)}{\pi^2}\frac{T^4}{\gamma}.
\ee
However, as is well-known \cite{Jeon:if} these one-loop results are not 
complete and a ladder summation is required to obtain the complete 
leading-order result (see Fig.\ \ref{figscalarladder}). Due to nearly 
pinching poles, each additional 
rung in the ladder contributes with a factor $\lambda^2 T/\gamma \sim 1$ 
and is therefore not suppressed. The effect of ladder summation is to 
change the coefficients $a_1, a_3,\ldots$, but not the parametric 
dependence on the coupling constant.

\begin{figure}[t]
\centerline{\epsfig{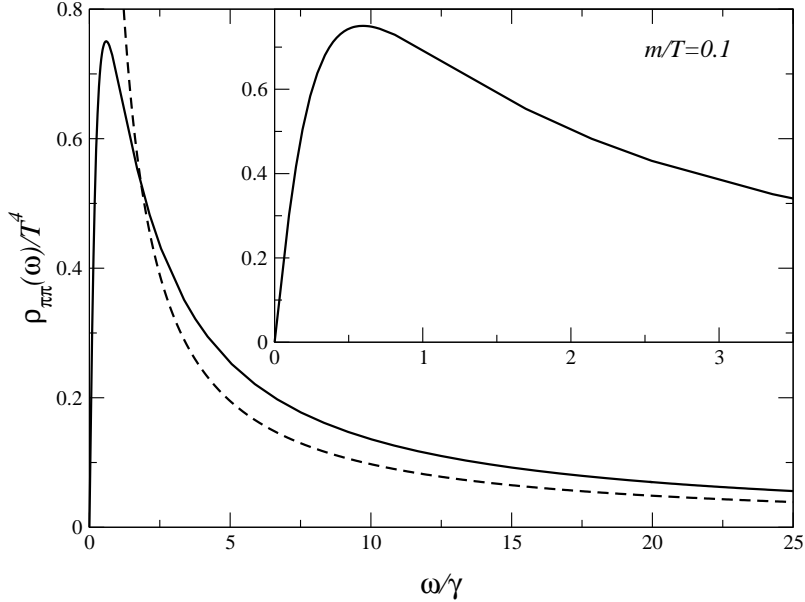}}
\caption{Contribution to the spectral function $\rho_{\pi\pi}(\om)/T^4$ 
(full line) from the nearly pinching poles in the low-frequency region as 
a function of $\om/\gamma$, obtained by numerical integration of Eq.\ 
(\ref{eqrhobelow}). The dashed line is the analytical result 
(\ref{eqapprox}) when $\gamma\ll\om\lesssim m$, neglecting nontrivial 
momentum dependence and finite mass corrections. The inset shows a blowup. 
The viscosity is determined by the slope for $\om\to 0$.
}
\label{figrhobelow}
\end{figure}

In the region $\gamma \ll \om \lesssim m$ we find
\be
\label{eqapprox}
\rho_{\pi\pi}(\om)/T^4 = -\frac{8}{3T^4} \int_\veck\, 
\frac{|\veck|^4}{\omk^2}
 n'(\omk) \frac{\gak}{\om} 
\simeq \frac{8\zeta(3)}{\pi^2}\frac{\gamma}{\om}
\;\;\;\;\;\;\;\;(\gamma \ll \om\lesssim m).
\ee
In this frequency interval pinching-pole contributions from ladders are
subdominant since each additional rung comes with a factor $\lambda^2 
T/\om \ll 1$. The perturbative part of the three-loop ladder diagram 
(i.e.\ with a single rung) also contributes at this order and has the same 
$\lambda^2T/\om$ behaviour as we find above.\footnote{We thank Guy Moore 
for pointing this out.}
Fig.\ \ref{figrhobelow} shows the
contribution to the spectral function from the nearly pinching poles in 
the low-frequency interval, obtained by numerical
integration of Eq.\ (\ref{eqrhobelow}) with the full momentum and mass
dependence. Note that the natural dimensionless combinations in the 
low-frequency region differ from those in the high-frequency region 
(compare Figs.\ \ref{figabove} and \ref{figrhobelow}). 

\begin{figure}[t]
\centerline{\epsfig{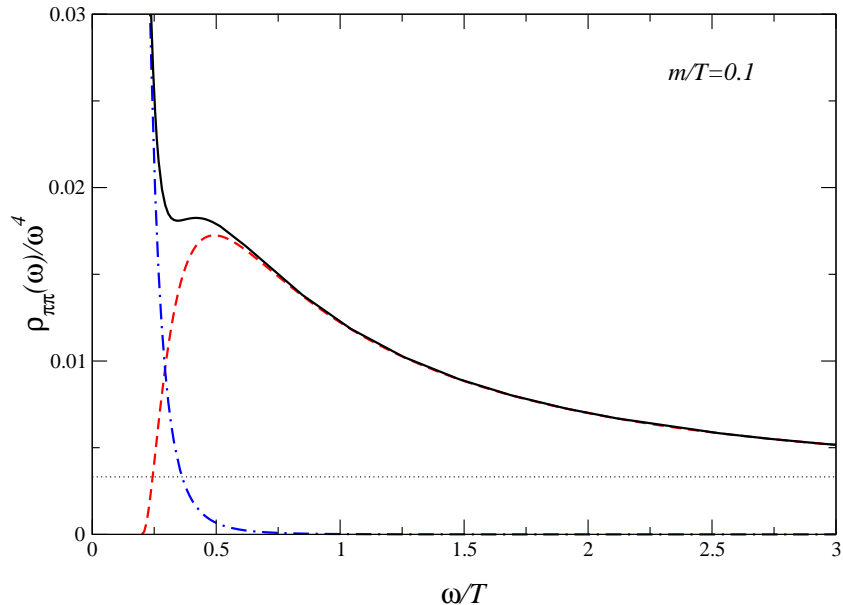}}
\caption{Complete one-loop spectral function $\rho_{\pi\pi}(\om)/\om^4$ 
(full line) as a function of $\om/T$, with $m/T=0.1$. The dashed line is 
the contribution from decay/creation processes and vanishes below $\om=2m$, 
the dot-dashed line is the contribution due to the nearly pinching poles 
at lower frequencies. The dotted line indicates the asymptotic value. 
}
\label{figrhoabbelow}
\end{figure}

We may now combine the results obtained so far to construct the complete 
one-loop spectral function at high temperature in the weak-coupling 
limit. The spectral function can be written as the sum of the 
contributions discussed above:
\be
\label{eqrholowhigh}
\rho_{\pi\pi}(\om) = \rho^{\rm low}_{\pi\pi}(\om) +
\rho^{\rm high}_{\pi\pi}(\om),
\ee
where $\rho^{\rm low}_{\pi\pi}$ represents the contribution from the
nearly pinching poles in Eq.\ (\ref{eqrhobelow}), dominating at low
frequencies, and $\rho^{\rm high}_{\pi\pi}$ is the contribution from 
decay/creation processes in Eq.\ (\ref{eqhigh}), dominating at higher frequencies.

We find that for very small frequencies the spectral function rises 
quickly as $\om/\gamma$. It reaches a maximum of order 1 (in units of 
temperature) at $\om\sim\gamma$ and decays then slowly as $\gamma/\om$.  
Around the thermal mass the contribution from decay/creation processes enter and 
the spectral function increases again. Note that the results obtained in 
both frequency domains smoothly match parametrically at $\om\sim m$, since
$\rho_{\pi\pi}(\om \sim 3m)/T^4 \sim \lambda^{3/2}$, both from the low- 
and the high-frequency calculation. For large $\om$ the spectral function
increases as $\om^4$, due to the zero-temperature decay process. Ladder 
diagrams do not affect this characteristic shape.
In Fig.\ \ref{figrhoabbelow} we present the complete one-loop spectral 
function as a function of $\om/T$.\footnote{ We used here that $m/T=0.1$ 
corresponds to $\lambda\simeq 0.267$ and $\gamma/T\simeq 1.48 \cdot 
10^{-5}$.}
In order to combine the low- and the high-frequency contribution in  
one figure, we show $\rho_{\pi\pi}(\om)/\om^4$, which enhances the 
contribution at lower frequencies.

We are now ready to calculate the euclidean correlator using 
Eq.\ (\ref{eqrel}). Because $G^{E}_{\pi\pi}(\tau)$ depends linearly on the 
spectral function, we write it as a sum of two contributions, 
$G^{E}_{\pi\pi} = G^{E,\rm low}_{\pi\pi} + G^{E,\rm high}_{\pi\pi}$, and 
discuss each term separately. We start with the contribution due to 
decay/creation processes which reads
\be
\label{eqGhigh}
G^{E,\rm high}_{\pi\pi}(\tau) = \int_{2m}^{\infty} \frac{d\om}{2\pi}\,
K(\tau,\om) 
\frac{\left(\om^2-4m^2\right)^{5/2}}{48\pi\om}\left[n(\om/2)+\half\right].
\ee
It is easy to see that the mass plays only a subdominant role and
finite-mass corrections are suppressed by $m^2/T^2$. Therefore we take  
$m=0$ which yields 
\be
\label{eqGm0}
G^{E,\rm high}_{\pi\pi}(\tau) =  \frac{T^5}{96\pi^2}\int_0^\infty 
dx\, 
x^4 \left[e^{sx} + 
e^{(1-s)x} \right] n(x) \left[n(x/2)+\half\right], 
\ee
where $x=\om/T$ and $s=\tau T$. The remaining integral can be performed 
and we find
\be
\label{eqGexact}
G^{E,\rm high}_{\pi\pi}(\tau) = \frac{\pi^2T^5}{3\sin^5 u}
\left\{ (\pi-u)\left[11\cos u+ \cos 3u\right] + 6\sin u+ 2\sin 3u
\right\},
\ee
where $u=2\pi \tau T$. An approximate but illuminating expression for the 
euclidean correlator can be obtained by noticing that the integral in Eq.\  
(\ref{eqGm0}) is dominated by hard frequencies, $x\gtrsim 1/s$ with
$0<s<1$, such that the Bose distributions may be approximated with 
Maxwellian ones, $n(x) \sim e^{-x}$. In that case we find 
\be
\label{eqana}
G^{E,\rm high}_{\pi\pi}(\tau) \simeq \frac{1}{8\pi^2}
\left[ 
\frac{1}{\tau^5} + \frac{1}{(1/T-\tau)^5} +
\frac{2}{(3/2T-\tau)^5} + \frac{2}{(1/2T+\tau)^5}
\right]. 
\ee
This approximate expression differs less than 2\% from the exact result 
(\ref{eqGexact}). The dominant $1/\tau^5$ behaviour of 
the correlator arises from decay at zero temperature. Finite temperature 
is manifested mainly through the reflection symmetry, $\tau \to 
1/T-\tau$. 
At the central point $\tau T=1/2$, we find the contribution to the 
euclidean correlator from the decay/creation processes to be
\be
G^{E,\rm high}_{\pi\pi}(\tau=1/2T) = 
\frac{4\pi^2}{45}T^5\left[1-\frac{25}{8\pi^2}\frac{m^2}{T^2} +\ldots 
\right].
\ee
The effect of a finite mass is to lower the mimimal value of the 
correlator.

\begin{figure}[t]
\centerline{\epsfig{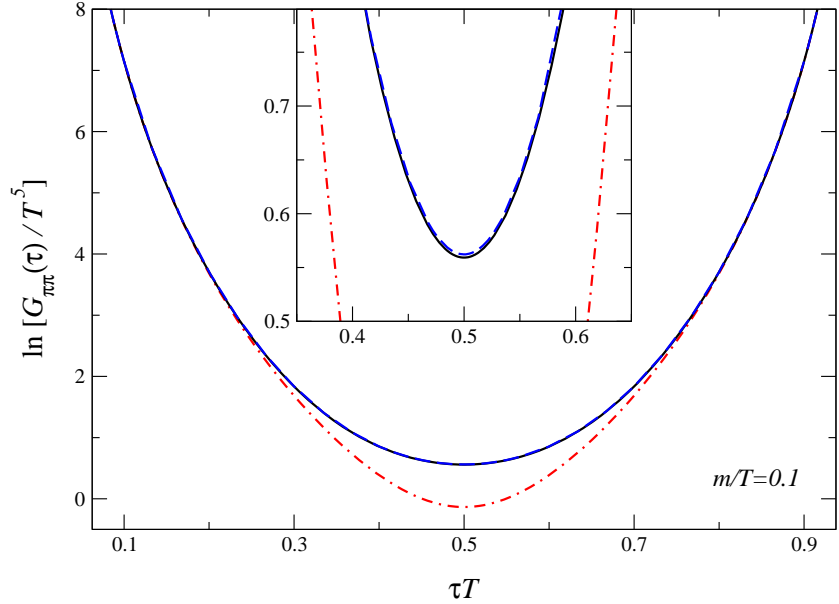}}
\caption{
Logarithm of the euclidean correlator $G_{\pi\pi}^E(\tau)/T^5$
as a function of $\tau T$, obtained by numerical integration of Eq.\ 
(\ref{eqrel}) with (\ref{eqrholowhigh}), for $m/T=0.1$ (full line).
The dashed line represents the analytical expression for $m=0$, Eq.\ 
(\ref{eqGEallexact}); it cannot be distinguished from 
the full result, except in the inset. The dot-dashed line shows the 
contribution from decay/creation processes only, obtained by numerical 
integration of Eq.\ (\ref{eqGhigh}). The inset shows a blowup around $\tau 
T=0.5$.
}
\label{figG}
\end{figure}

The contribution to the euclidean correlator from the nearly pinching 
poles in the low-frequency region can be found easily from Eqs.\ 
(\ref{eqrel}) and (\ref{eqrhobelow}) by interchanging frequency and 
momentum integrals. Since $\rho_{\pi\pi}^{\rm low}$ gives the dominant 
contribution to the spectral function up to frequencies of order $m$, 
whereas $\rho_{\pi\pi}^{\rm high}$ dominates for higher frequencies, we 
can use the expansion (\ref{eqexp}) for the kernel and obtain 
\bea
\nn
G^{E,\rm low}_{\pi\pi}(\tau)  &\simeq&  -\frac{8}{3} \int_\veck\, 
\frac{|\veck|^4}{\omk^2} n'(\omk)
\int_0^{\om_{\Lambda}} \frac{d\om}{2\pi}\, 
\frac{2T}{\om}\frac{\om\gak}{\om^2+4\gak^2}\\
\nn
&=&
-\frac{4}{3}\frac{T}{\pi} \int_\veck\, 
\frac{|\veck|^4}{\omk^2} n'(\omk) \arctan 
\left(\frac{\om_{\Lambda}}{2\gak}\right)\\
&\simeq& \frac{4\pi^2}{45}T^5 \left[ 1 -\frac{25}{8\pi^2}\frac{m^2}{T^2} 
+\ldots\right],
\eea
with $\om_{\Lambda}\sim m$. The error that is introduced by expanding the
kernel is negligible (see Fig.~\ref{figG}). We note that in our one-loop
calculation this result is at leading order independent of the coupling
constant. Similarly, while ladder diagrams determine the precise shape of 
the spectral function, the effect on the euclidean correlator $G^{E,\rm
low}_{\pi\pi}(\tau)$ appears only in subleading corrections in the
coupling constant, as can be seen with the kinetic approach \cite{Guy}. 
Therefore the low-frequency contribution to the euclidean correlator is 
constant and of order one (in the appropriate units) and insensitive to 
details of the ladder summation.

Combining the low- and high-frequency contribution to the euclidean
correlator at high temperature and weak coupling we find, to leading order 
in the coupling constant, 
\be
\label{eqGEallexact}
G^E_{\pi\pi}(\tau) = 
\frac{\pi^2T^5}{3\sin^5 u}
\left\{ (\pi-u)\left[11\cos u+ \cos 3u\right] + 6\sin u+ 2\sin 3u
\right\} + \frac{4\pi^2}{45} T^5,
\ee
%\hm&&\simeq \frac{1}{8\pi^2}
%\left[ 
%\frac{1}{\tau^5} + \frac{1}{(1/T-\tau)^5} +
%\frac{2}{(3/2T-\tau)^5} + \frac{2}{(1/2T+\tau)^5}
%\right]
%+ \frac{4\pi^2}{45} T^5,
%\eea
with $u=2\pi \tau T$.
Corrections due to the finite thermal mass are
suppressed by $m^{2}/T^{2}$. The euclidean correlator is minimal at $\tau
T=1/2$, and $G_{\pi\pi}^{E}(1/2T)$ receives contributions of the same
order from both the high- and the low-frequency region (at leading 
order in the coupling they are equal). 
A comparison between the analytical result at $m=0$ and the full result 
obtained by numerical integration of Eq.\
(\ref{eqrel}) with Eq.\ (\ref{eqrholowhigh}) in the presence of a finite
mass is shown in Fig.\ \ref{figG}.

\section{Nonabelian gauge fields}
\label{secgauge}

\begin{figure}
\centerline{\epsfig{figure=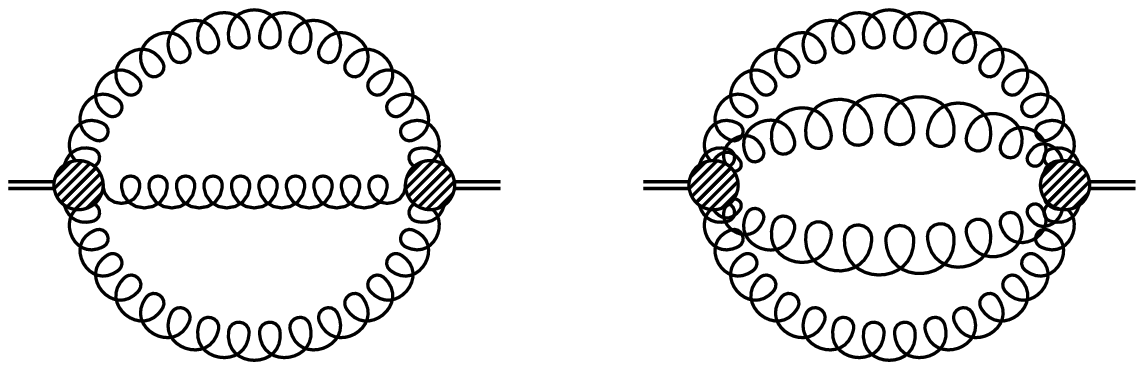,height=3.2cm}}
\caption{Diagrams that contribute to the spectral function 
$\rho_{\pi\pi}$ at higher order. These diagrams are special for a 
nonabelian theory. 
}
\label{figgluons34}
\end{figure}

We leave the scalar case and consider nonabelian $SU(N_c)$ gauge 
theory. The traceless spatial part of the energy-momentum tensor is
\be
\label{eqpi}
 \pi_{ij} = F_{i}^{a\mu}F_{j\mu}^{a} 
 - \frac{1}{3}\delta_{ij}F^{ak\mu}F_{k\mu}^{a}.
\ee
The coupling vertex between the operator $\pi_{ij}$ and two
gluons with incoming momenta $P,K$ and indices $(\mu,a),(\nu,b)$
respectively can be read from (\ref{eqpi}) and we find\footnote{
In order to arrive at the basic one-loop expression (\ref{one-loop}) 
below we use here the imaginary-time formalism with $P=(p_4,\pv)$, 
the Matsubara frequency $\om_n = -p_4 = 2\pi nT$ ($n\in \mathbb{Z})$ 
and $P\cdot K=p_4k_4+\pv\cdot\kv$.}
\bea
 \nn
 \Gamma^{ab}_{ij,\mu\nu}(P,K) \!\!\!&=&\!\!\! -\delta^{ab}
 \Bigg[ 
 \delta_{\mu\nu} 
 \left( p_{i}k_{j} + p_{j}k_{i} - \frac{2}{3}\delta_{ij}\pv\cdot\kv \right) 
 \\ \nn
 &&\!\!\! + P\cdot K \Big( \delta_{i\mu}\delta_{j\nu}
 + \delta_{i\nu}\delta_{j\mu} 
 - \frac{2}{3}\delta_{ij}\delta_{k\mu}\delta_{k\nu} \Big) 
 - ( p_{i}K_{\mu}\delta_{j\nu} 
 + p_{j}K_{\mu}\delta_{i\nu} 
 \\  &&\!\!\!
 + P_{\nu}k_{i}\delta_{j\mu} 
 + P_{\nu}k_{j}\delta_{i\mu} )
 + \frac{2}{3}\delta_{ij}\left( P_{\nu}k_{k}\delta_{k\mu} 
 + p_{k}K_{\mu}\delta_{k\nu} \right)
 \Bigg].
\eea
In the nonabelian theory $\pi_{ij}$ also couples to three and four 
gluons, which leads to diagrams as depicted in Fig.\ \ref{figgluons34}. 
However, these contributions are suppressed by powers of the coupling 
constant and will not be considered further. 

We use the Coloumb gauge in which the gluon propagator reads
\be
 D_{\mu\nu}(P) = \cP^{T}_{\mu\nu}(\pvuni)\Delta_{T}(P) 
               + \delta_{4\mu}\delta_{4\nu}\Delta_{L}(P), 
\ee
with $\cP^{T}_{ij}(\pvuni) = \delta_{ij}-\puni_{i}\puni_{j}$, 
$\cP^{T}_{4\mu} = \cP^{T}_{\mu4} = 0$. The transverse and
longitudinal components have the following spectral representations
\be   
 \Delta_{T}(P) = - \int_{-\infty}^{\infty}\frac{d\omega}{2\pi}\, 
 \frac{\rho_{T}(\omega,\pv)}{i\omega_n-\omega},
 \;\;\;\;\;\;
 \Delta_{L}(P) = \frac{1}{\pv^2} + 
 \int_{-\infty}^{\infty} \frac{d\omega}{2\pi}\, 
 \frac{\rho_{L}(\omega,\pv)}{i\omega_n-\omega}. 
\ee
The one-loop contribution to the spectral function reads, after 
evaluating the Matsubara sum and taking all frequencies real again, 
\bea
\nn 
 \rho_{\pi\pi}(\om) = &&\hm \frac{2d_{A}}{3} \int\frac{d^4k}{(2\pi)^4}
 \left[n(k^{0})-n(k^{0}+\qo)\right]
 \Bigg\{
 V_{1}(k,\om)\rho_{T}(k^0,\veck)\rho_{T}(k^0+\om,\veck)
 \\ && \hm
 +  V_{2}(k) \rho_{T}(k^0,\veck)\rho_{L}(k^0+\om,\veck)
 + V_{3}(\veck) \rho_{L}(k^0,\veck)\rho_{L}(k^0+\om,\veck)
 \Bigg\},
\label{one-loop}
\eea
where $d_{A}=N_{c}^{2}-1$ is the number of gluons and $V_1(k,\om) = 
7|\veck|^4 - 10\veck^2 k^0(k^0+\om) + 7(k^0)^{2}(k^0+\om)^{2}$, 
$V_{2}(k) = 6\veck^2(k^0)^2$ and  $V_{3}(\veck) =  - 32 |\veck|^4$. 
This expression is the equivalent of Eq.\ (\ref{eqrho}) in the scalar
theory. 

As in the scalar case, we start the analysis for external frequencies 
$\om$ that are sufficiently large such that no pinch singularities are 
present. The collective (HTL) effects in a nonabelian plasma 
\cite{Braaten:1989mz} can be 
incorporated in the one-particle spectral functions, which are however 
more complex than in the scalar case. The spectral function for transverse 
gluons can be written as \cite{LeBellac,Blaizot:2001nr}
\be
 \rho_{T}(k^{0},\veck) = 2\pi Z_{T}(|\veck|) 
 \left\{ \delta[k^{0}-\omega_{T}(|\veck|)] - 
 \delta[k^{0}+\omega_{T}(|\veck|)]\right\}
 + \beta_{T}(k^0, |\veck|).
\ee
The delta functions describe propagating quasiparticles with a 
dispersion relation $\om_T$ and a residue 
\be
 Z_{T}(k) = \frac{ \omega_T(k) [\omega_T^2(k) - 
 k^{2}]}{3\wp^2\omega_T^2(k) - [\omega_T^2(k)-k^2]^2},
\ee
where $\wp^2 = g^{2}T^{2}N_{c}/9$ is the plasma frequency squared. 
For small and large spatial momentum one finds 
\bea
 &&\om_T^2(k) \simeq \wp^2 +\frac{6}{5} k^2, \;\;\;\;\;\;\;\;
 Z_{T}(k) \simeq 1/(2\wp) \;\;\;\;\;\;\;\; (k\to 0), \\
 &&\om_T^2(k) \simeq k^2+m^2_\infty, \;\;\;\;\;\;\;\;\;
 Z_{T}(k) \simeq 1/(2k) \;\;\;\;\;\;\;\;\;\; (k\to \infty),
\eea
where $m^2_\infty = \frac{3}{2}\wp^2$ is the asymptotic gluon mass 
squared. The $\beta_T$ function describes Landau damping and is 
nonzero below the light-cone only.
The spectral function of longitudinal gluons $\rho_{L}(k^{0},\veck)$ has a
similar form \cite{LeBellac,Blaizot:2001nr}.

We start studying the contribution to Eq.\ (\ref{one-loop}) when both 
gluons are transverse. There are three parts, depending on whether we take 
from the one-particle spectral functions the delta functions of the 
quasiparticles (pole-contribution) or the $\beta$ function of the Landau 
damping  (cut-contribution). When both gluons are quasiparticles the 
integrals can be done with the help of the delta functions and we 
find the pole-pole contribution for transverse gluons
\be
 \label{eqttpp}
 \rho^{\rm pp}_{\pi\pi}(\om) = 
 \theta(\om-2\wp)\frac{2d_{A}}{3\pi}
 \frac{Z_T^2(f_0)  f_{0}^{2}}{\omega'_{T}(f_{0})}
 \left( 7f_0^4 + \frac{5}{2}f_0^2\om^{2} + \frac{7}{16}\om^{4} \right)
 \left[n(\om/2)+\frac{1}{2}\right].
\ee
Here we use the notation $f_{0}=f(\om/2)$, with $f(u)$ defined as the 
inverse of the transverse dispersion relation, $f[\om_T(k)] = k$. 
The function $f(u)$ vanishes when $u\leq \wp$, and
\bea
 &&f(u) \simeq \sqrt{ \frac{5}{6} (u^2-\wp^2) }  
 \;\;\;\;\;\;\;\;\;\;\;\; (u\to \wp),\\
 &&f(u) \simeq u - \frac{3\wp^2}{4u} 
 \;\;\;\;\;\;\;\;\;\;\;\;\;\;\;\;\;\;\;\; (u\to \infty).
\eea
Again, as for the scalar case, the HTL resummation produces the threshold 
in the spectral function (\ref{eqttpp}) for soft external frequency. 
We find that 
\be
 \label{eqttppsoft}
 \rho^{\rm pp}_{\pi\pi}(\om\sim gT) \sim g^3 T^4.
\ee
For large $\om$ the spectral function behaves as 
\be 
 \label{high}
 \rho^{\rm pp}_{\pi\pi}(\om) = \frac{d_A}{4\pi} 
\om^{4}\left[n(\om/2)+\frac{1}{2}\right] \;\;\;\;\;\;\;\;\;\;\;\;\;
(\om\gg\wp),
\ee
which is, up to the prefactor, what we found in the scalar case as well.

The contribution when one transverse gluon is a quasiparticle and the 
other undergoes Landau damping (pole-cut contribution) is
\bea  
 \label{p-c}
 \rho^{\rm pc}_{\pi\pi}(\om) =  \frac{2d_A}{3\pi^2} 
 \int_{\wp}^{\infty}du &&\hm \frac{f^{2}(u)}{\omega'_{T}[f(u)]}Z_{T}[f(u)]
 [n(u-\om)-n(u)] \beta_{T}[u-\om,f(u)] 
 \nonumber \\
 &&\hm \times \left[ 7f^4(u) - 10u(u-\om)f^2(u) + 7u^2(u-\om)^2 \right].
\eea
For soft $\qo\sim gT$ the dominant contribution arises when the energy $u$ 
is hard, and we may use 
\be
 \beta_T[u-\om,f(u)] \sim \frac{3\pi}{4}\frac{\wp^2}{\om u^3},
\ee
as well as other simplifications given above. We find
\be
 \label{eqpc}
 \rho^{\rm pc}_{\pi\pi}(\om \sim gT) \simeq  
 -\frac{d_A}{\pi} \wp^2  
 \int_{\wp}^{\infty} du\, u^2 n'(u) \sim g^{2}T^{4}.
\ee
For soft external frequencies the pole-cut contribution 
dominates over the pole-pole contribution (\ref{eqttppsoft}). 

For hard frequencies $\om\sim T$, the $\beta_T$ function determines the 
lower integration limit to be $u_0 = \om/2+ 3\wp^{2}/(4\om)$. As a 
result, $u$ is always hard and we can simplify the integrand to arrive at 
\be
 \rho^{\rm pc}_{\pi\pi}(\om \sim T) =  \frac{d_A}{3\pi^2} 
 \int_{u_0}^{\infty} \! du\, u^3
 \left( 4u^2 - 4u\om + 7\om^2 \right) [n(u-\om)-n(u)] 
 \beta_{T}(u-\om, u).
\ee
It is convenient to substitute $u = \om (z+1)/2$ such that
\be
 \rho^{\rm pc}_{\pi\pi}(\om \sim T) =  \frac{d_A\om^6}{48\pi^2}
 \int_{a}^{\infty} \! dz\, (z+1)^3
 \left( z^2+6 \right) [n(\om z_-)-n(\om z_+)] 
 \beta_{T}(\om z_-, \om z_+),
\ee
where $z_{\pm} = (z\pm 1)/2$ and $a = 3\wp^{2}/(2\om^2)$. The dominant 
contribution comes from the lower integration limit ($z\to a$) 
where the $\beta_{T}$ function can be approximated as 
\be
 \beta_{T}(\om z_-, \om z_+) \simeq 
 -\frac{4\pi}{\om^{2}}\frac{z a}{(z+a)^2}.
\ee
Using this expression we find that 
\be
 \rho^{\rm pc}_{\pi\pi}(\om \sim T) \sim g^{2}T^{4}\ln(1/g).
\ee 
For hard frequencies the pole-cut contribution is therefore suppressed 
compared to the pole-pole contribution (\ref{high}). 
Finally, we found that the remaining cut-cut contribution when both 
gluons are transverse is suppressed with respect to the pole-pole 
contribution when $\om$ is hard and to the pole-cut contribution when 
$\om$ is soft. 

In a similar way we have analysed the remaining longitudinal-transverse
and longitudinal-longitudinal contributions in Eq.~(\ref{one-loop}) with
the result that they do not modify the conclusions drawn from the
transverse-transverse contribution analysed above. In particular, for hard
frequencies $\om\gg\wp$ transverse gluons dominate and the spectral
function is given by Eq.\ (\ref{high}). We do not need to be more explicit
about the region where $\qo \siml gT$ because the dominant contribution
in this region arises from the pinching poles, screened by a finite width
and/or external frequency, as we will show now.

For small external frequencies $0 \leq \om \lesssim gT$ the loop 
integral is dominated by the region of hard momentum and we only need to 
consider transverse gluons, since the residue for longitudinal gluons 
vanishes exponentially. In order to avoid pinch singularities we follow 
the same steps as in the scalar case and substitute for the one-particle 
spectral function a Breit-Wigner function, see Eq.\ (\ref{eqBW}). For 
transverse on-shell gluons with hard momentum the dispersion relation is 
$\omega_{\veck} = \sqrt{\veck^2 + m_{\infty}^{2}}$ and 
the leading (momentum-independent) contribution to the damping rate 
is~\cite{Pisarski:rf}
\be
 \gamma=\frac{g^{2}}{4\pi}N_{c} T \ln(1/g),
\ee
where the logarithm is sensitive to the magnetic mass, $\ln(\wp/m_{\rm 
mag})\sim \ln(1/g)$ with $m_{\rm mag}\sim g^{2}T$.\footnote{In 
QED there is no magnetic mass which could regularize the logarithmic 
divergence of the leading contribution to the damping rate. It 
turns out that the electron damping rate is ill 
defined~\cite{Blaizot:1996az}.}
The mass $m_\infty$ only plays a subdominant role and is neglected below. 
Evaluating the integral over $k^{0}$ exactly as in the scalar case, we 
find 
\bea
 \rho_{\pi\pi}(\om) = -\frac{d_A}{3} \int_\veck\, 
 \frac{1}{|\veck|}
 \Big\{ &&\hm
 [n(|\veck|)-n(|\veck|-\om)] 
 I(\om,\veck) A(\om,\veck) \frac{ |\veck|-\om}{\om^2+4\gamma^2}
 \nn \\  &&\hm
 -[\om\to-\om] \Big\},
\eea
where $I(\om,\veck)$ was defined in Eq.\ (\ref{eqI}) and
$A(\om,\veck) = \veck^2\left( 4\veck^2- 4\om|\veck| + 7\omega^{2}\right)$.
For $\om \lesssim \wp$ this expression simplifies to  
\be
\label{eqgauge}
\rho_{\pi\pi}(\om) = -\frac{16d_A}{3} \int_\veck\, \veck^2
 n'(|\veck|) \frac{\om\gamma}{\om^2+4\gamma^2} 
\;\;\;\;\;\;\;\;(0\leq\om \lesssim \wp),
\ee
which is $2d_A$ times the scalar result. 
In the region $\gamma\ll\qo\siml\wp$ we find
\be
 \label{eqdown}
 \rho_{\pi\pi}(\qo)/T^4 = -\frac{16d_A}{3T^4}  
	\frac{\gamma}{\qo} \int_\veck\,  \veck^2 n'(|\veck|)
 = \frac{32\pi^2}{45}d_A \frac{\gamma}{\qo}.
\ee
As in the scalar case, the three-loop ladder diagram with one rung 
contributes in this region at the same order. Note that the contribution 
from the pinching poles at $\om \sim gT$, 
\be
 \label{eqBWgauge}
 \rho_{\pi\pi}(\om\sim gT) \sim gT^4 \ln (1/g)
\ee
actually dominates over the contribution (\ref{eqpc}) from the collective 
(HTL) excitations in this region.  
The shear viscosity in the one-loop approximation follows from 
(\ref{eqgauge}) as
\be
\label{eqone-loopsh}
\eta_{\rm 1-loop} = \frac{8\pi^2 d_A}{225}\frac{T^4}{\gamma}.
\ee
However, for the complete spectral function at small frequencies 
$0\leq\qo\siml\gamma$ the effects of ladder diagrams must be taken into 
account. As was mentioned in the Introduction, 
kinetic theory predicts that the shear viscosity is parametrically larger 
than the one-loop result (\ref{eqone-loopsh}) 
\cite{Pisarski:1992hy,Arnold:2000dr,Baym:uj}. 
Therefore, for vanishing frequency the spectral function due to ladder 
diagrams is expected to be larger than the one-loop contribution 
and to behave as 
$\rho^{\rm ladder}_{\pi\pi}(\qo\to 0) = 20 \eta\, \om$, 
with \cite{Arnold:2000dr}
\be
\eta \sim \frac{N_c^2-1}{N_c^2} \frac{T^3}{g^{4}\ln(1/g)}.
\ee
Hence the slope of the spectral function is much steeper close to the
origin, compared to the one-loop result. Up to frequencies $\qo\sim\gamma$
the actual behaviour of the spectral function depends on the 
pinching-poles contribution of ladders diagrams. When
$\gamma\ll\om \lesssim \wp$ pinching-poles contributions of the ladders 
are subdominant by a factor $g^2 T/\qo$ and the spectral function 
decreases as $g^2T/\om$ until it meets with the rising contribution from 
Eq.\ (\ref{high}). For large $\om$ the spectral function increases as 
$\om^4$.

The euclidean correlator at high temperature can be computed as in the 
scalar case and we find, at leading order in $g$, 
\be
G^E_{\pi\pi}(\tau)/T^5 = 
\frac{4\pi^2d_A}{\sin^5 u}
\left\{ (\pi-u)\left[11\cos u+ \cos 3u\right] + 6\sin u+ 2\sin 3u
\right\} + \frac{8\pi^2d_A}{45},
\ee
with $u=2\pi \tau T$. The last constant term reflects the low-frequency 
region.

\section{Conclusions}

We have studied the spectral function relevant for the shear viscosity in
scalar and nonabelian gauge theories at high temperature as a function of
the external frequency. While for small frequencies ladder diagrams are
important in the scalar case and essential in the nonabelian case, for
higher frequencies a simple one-loop computation yields the dominant
contribution.

We found that the spectral function has a characteristic shape: for small
frequencies the spectral function rises, reaches a local maximum and
decreases as $1/\om$.  This contribution is due to scattering processes
in the plasma and is enhanced due to nearly pinching poles. We referred
to this contribution as the low-frequency contribution.  For higher
frequencies, decay/creation processes dominate and the spectral function 
increases essentially as $\om^4$. This contribution is referred to as the
high-frequency contribution.

In order to extract transport coefficients from euclidean lattice
correlators, a simple ansatz, essentially a Breit-Wigner spectral
function, was introduced in Ref.\ \cite{Karsch:1986cq} to model spectral
functions of components of the energy-momentum tensor. The resulting
three-parameter ansatz was subsequently used in Refs.\
\cite{Nakamura:1996na,Nakamura:1997bh} to determine transport
coefficients in hot gauge theories from lattice simulations.
Unfortunately, as we have seen in this paper, at high temperature
spectral functions of composite operators do not resemble simple
Breit-Wigner functions at all. In order to improve this analysis, we 
propose therefore to use a different ansatz, which is written as the sum 
of two terms:
\be
\label{eqansatz}
\rho_{\pi\pi}(\om) = \rho^{\rm low}_{\pi\pi}(\om) + 
\rho^{\rm high}_{\pi\pi}(\om).
\ee 
The high-frequency part can be described 
by Eqs.\ (\ref{high}) or (\ref{eqhigh}) with $m$ as a possible free 
parameter. For the low-frequency 
part we note that the spectral function is odd, increases linearly with 
$\om$ for small $\om$ and decreases with $1/\om$ for larger 
$\om$. A simple ansatz reflecting this is
\be
\rho^{\rm low}_{\pi\pi}(x)/T^4 = 
x\frac{b_1+b_2x^2+b_3x^4+\ldots}{1+c_1x^2+c_2x^4+c_3x^6+\ldots}, 
\;\;\;\;\;\;\;\;x=\om/T,
\ee 
with $b_i=c_i=0$, $i> n$ for given $n$.  The viscosity is given by
$\eta/T^3 = b_1/20$. 
The ansatz (\ref{eqansatz}), with free parameters $m$, $b_i$, and $c_i$, 
should be used in Eq.\ (\ref{eqrel}) to fit the corresponding euclidean 
correlator to the numerical results. 
When insisting on a three-parameter fit, one may take
$n=1$ which leaves $m$, $b_1$ and $c_1$ to be determined.

Concerning the euclidean correlator, we found that the dominant $\tau$
dependence is determined by the high-frequency part ($\om \gtrsim T$) of
the spectral function. However, around $\tau T=1/2$ both the high- and the
low-frequency regions of the spectral function contribute at the same
order. The low-frequency contribution is of special importance since
transport coefficients are determined by the slope of the spectral
function at zero frequency and a precise calculation of the spectral
function at low frequencies is therefore essential.  We found that no
matter how complicated the spectral function up to frequencies $\om\sim g
T$ ($g \ll 1$) might be, its contribution to the euclidean correlator will 
be a $\tau$ independent constant. This latter feature poses a
severe challenge for the MEM approach, since the spectral function at low
frequencies should be reconstructed from the knowledge of a single
constant alone. It turns out that this constant carries information on the 
transport coefficient (or more generally on ladder diagrams) only in 
subleading contributions. As a result euclidean correlators are  
remarkably insensitive to transport coefficients, which makes it extremely 
difficult to extract those at weak coupling. 

We emphasize that these results are not specific for the correlator we
considered here. For instance, in the first paper of Ref.\
\cite{Karsch:2001uw} the current-current correlator relevant for thermal
dilepton rates in QCD was studied on the lattice and an enhancement in the
central value of the euclidean correlator compared to the free result was
observed. This enhancement might be accounted for by the pinching-poles
contribution in the low-frequency region of the spectral
function.\footnote{Note, however, that in the lattice study the spectral
function reconstructed with the Maximal Entropy Method seems to vanish
below $\om\sim 3 T$.}

\vspace{0.5cm}
\noindent
{\bf Acknowledgements}\\
It is a pleasure to thank Ulrich Heinz and Eric Braaten for discussions 
and comments. G.~A.\ thanks Jan Smit and Nucu Stamatescu as well.
We thank Guy Moore for a careful reading of and valuable comments on the 
manuscript. 
G.~A.\ is supported by the Ohio State University through a Postdoctoral
Fellowship and by the U.~S.\ Department of Energy under Contract No.\
DE-FG02-01ER41190. 
J.~M.~M.~R. is supported by a Postdoctoral Fellowship from the Basque 
Government.

\end{document}